\documentclass[a4paper,12pt]{article}
\pdfoutput=1 % if your are submitting a pdflatex (i.e. if you have
\usepackage{jheppub} % for details on the use of the package, please
\usepackage{hyperref}
\hypersetup{
    bookmarks=true,         % show bookmarks bar?
    unicode=false,          % non-Latin characters in AcrobatÃÂÃÂÃÂÃÂ¢ÃÂÃÂÃÂÃÂÃÂÃÂÃÂÃÂs bookmarks
    pdftoolbar=true,        % show AcrobatÃÂÃÂÃÂÃÂ¢ÃÂÃÂÃÂÃÂÃÂÃÂÃÂÃÂs toolbar?
    pdfmenubar=true,        % show AcrobatÃÂÃÂÃÂÃÂ¢ÃÂÃÂÃÂÃÂÃÂÃÂÃÂÃÂs menu?
    pdffitwindow=false,     % window fit to page when opened
    pdfstartview={FitH},    % fits the width of the page to the window
    pdftitle={My title},    % title
    pdfauthor={Author},     % author
    pdfsubject={Subject},   % subject of the document
    pdfcreator={Creator},   % creator of the document
    pdfproducer={Producer}, % producer of the document
    pdfkeywords={keyword1} {key2} {key3}, % list of keywords
    pdfnewwindow=true,      % links in new window
    colorlinks=true,       % false: boxed links; true: colored links
    linkcolor=red,          % color of internal links
    citecolor=purple,        % color of links to bibliography
    filecolor=magenta,      % color of file links
    urlcolor=blue,           % color of external links
    linktocpage=true
}
\usepackage{graphics}
\usepackage{rotating}
\usepackage{subfigure}
\usepackage{pstricks}
\usepackage{color}
\usepackage{amsfonts}
\usepackage{mathrsfs}
\usepackage{epsfig}
\usepackage{amsmath,amssymb,amsthm,graphicx,latexsym}
\usepackage{ulem}

\newcommand{\be}{\begin{equation}}
\newcommand{\ee}{\end{equation}}
\newcommand{\bea}{\begin{eqnarray}}
\newcommand{\eea}{\end{eqnarray}}
\newcommand{\bml}{\begin{subequations}}
\newcommand{\eml}{\end{subequations}}
\newcommand{\bfig}{\begin{figure}}
\newcommand{\efig}{\end{figure}}

\newcommand{\slashed}{\hspace{-1.1ex}/}

\begin{document}
$~~~~~~~~~~~~~~~~~~~~~~~~~~~~~~~~~~~~~~~~~~~~~~~~~~~~~~~~~~~~~~~~~~~~~~~~~~~~~~~~~~~~$\textcolor{red}{\large\bf TIFR/TH/15-09}
\title{\textsc{\fontsize{45}{90}\selectfont \sffamily \bfseries  \textcolor{purple}{Fermion localization in higher curvature spacetime}
		}}

\author[a]{Sayantan Choudhury
\footnote{\textcolor{purple}{\bf Presently working as a Visiting (Post-Doctoral) fellow at DTP, TIFR, Mumbai, \\$~~~~~$Alternative
 E-mail: sayanphysicsisi@gmail.com}. ${}^{}$},}
\author[b]{Joydip Mitra}
\author[c]{and Soumitra SenGupta}

\affiliation[a]{Department of Theoretical Physics, Tata Institute of Fundamental Research, Colaba, Mumbai - 400005, India
}
\affiliation[b]{Department of Physics, Scottish Church College, 1 \& 3 Urquhart Square
Kolkata - 700 006, India
}
\affiliation[c]{Department of Theoretical Physics,
Indian Association for the Cultivation of Science,
2A \& 2B Raja S. C. Mullick Road,
Kolkata - 700 032, India.
}

\emailAdd{sayantan@theory.tifr.res.in, jmphys@scottishchurch.ac.in, tpssg@iacs.res.in  }

\abstract{Fermion localization in a braneworld model in presence of dilaton coupled higher curvature Gauss-Bonnet bulk gravity is discussed.
It is shown that the lowest mode of left handed fermions can be naturally localized on the visible brane due to the dilaton coupled higher curvature term without the necessity of any external localizing 
bulk field. 
}

%\begin{document} 
\maketitle
\flushbottom
\section{Introduction}
\label{aa1}
Randall Sundrum warped geometry model \cite{Randall:1999ee,Randall:1999vf} is an eminently successul model in resolving the long standing gauge hierarchy/naturalness problem
in an otherwise successful Standard model of elementary particles. This resulted
in extensive search for a signature of Randall Sundrum (RS) model in Large Hadron Collider (LHC) \cite{Weiglein:2004hn,Skittrall:2008pr,DePree:2006ah,
Aad:2012cy,Tang:2012pv,Das:2013lqa} . When applied as a physics beyond Standard Model (SM) such a scenario is often based on an underlying assumption
that the SM fermions in general can propagate in the bulk while their chiral states are appropriately localized in different regions of bulk spacetime
producing the desired 4-dimensional fermion masses on the visible brane.
Though RS model itself can not provide any justification for this localization, there have been efforts to provide
an explanation of this by introducing an ad-hoc scalar field in the anti- de Sitter bulk of RS model
\cite{Bajc:1999mh,Lehners:2007xa,Chang:1999nh,Koley:2004at,ArkaniHamed:1999za,Kaplan:2001ga,Grossman:1999ra}.
Such a mechanism of localization however gives rise to the speculation about the possible back-reaction of the scalar to jeopardize the original RS solution 
for the warp factor.
The origin of hierarchy of fermion masses in Standard model is a problem yet to be resolved. In this work
we show that a string inspired modification to Einstein gravity via dilaton and higher curvature effects
can explain the hierarchy among the fermion masses, where the effective masses of fermions are determined
by their couplings with the dilaton field. Our work is staged on a five dimensional spacetime where our 
universe sits at one of the fixed points of the orbifolded extra spacetime dimension.
This work thus is an attempt to look for an alternative pathway for this localization which resorts to the presence
of higher curvature corrections to the classical Einsteinian gravity in the bulk as postulated in RS model. Such a correction, though 
heavily suppressed in low energy world, assumes significance in an Anti de Sitter (AdS) bulk with curvature $\sim$
Planck scale as assumed in RS model. At the leading order, such correction over Einstein Gravity which is free from the appearance of any 
ghost field due to the higher derivative terms, is the Gauss-Bonnet (GB) gravity
where the various quadratic curvature terms appear in suitable combination to make the theory stable.
Such a correction is also inspired by string theory which in addition  also predicts the presence of the scalar dilaton in the action.

Following the path adopted in RS model, if the  dilaton coupled GB gravity is compactified on $S^1/Z_2$ orbifold,  it gives
rise to two branches of warped solutions, one of which is stable and free of ghost. In this work we adopt
this ghost free branch and study the effects of higher curvature couplings on the two chirality states
of 5 dimensional SM fermions to study their localization on the standard model 3-brane.  

As discussed earlier, if the standard model fermions also propagate in the bulk, just as graviton, then  
by appropriately choosing  the interaction potential between the fermion and a 
localizing field one can try to obtain the appropriate overlap between the chiral states of the fermionic wave functions to localize the 
left chiral massless fermion states on the brane.  However without resorting to any such external ad-hoc field can one 
produce this desired feature through the higher curvature GB terms in the bulk?
This would then provide a natural explanation of observing only the left-handed neutrinos in our universe while the massive fermions appear with 
both the chiral states.    
We try to address this question in the present work in the framework of GB-dilaton induced warped geometry model \cite{Choudhury:2013yg}.

In this work, we consider a bulk fermion and study the localization profile in the five dimensional warped geometry model
in the backdrop of GB dilaton gravity. We will show that inclusion of higher curvature terms lead to the localization of left handed fermionic modes
near the visible brane as their mass decreases while the right hand modes are localized within the bulk. In such localization scenario therefore one does not need to 
invoke an external bulk field as has been proposed earlier. Since the Standard model only includes 
the left handed modes of
massless fermions, we can clearly see that this higher curvature setup will automatically allow the fermionic wave functions to localize themselves in the TeV brane.
On the other hand, the delocalization of right handed modes inside the bulk can produce interesting phenomenological consequences like 
fermion mass generation as suggested by Ref.~\cite{Grossman:1999ra}.
  
\section{Fermion localization in Gauss-Bonnet dilaton gravity}
Here we generalize the analysis in warped geometry
 in presence of Gauss-Bonnet coupling and gravidilaton coupling in a 5D bulk.
 The background warped geometry model
is proposed by making use of the following sets of assumptions as a building block:
\begin{itemize}
 \item The leading order Einstein's gravity sector is modified by the GB 
\cite{Choudhury:2012yh,Choudhury:2013qza,Choudhury:2013eoa,Choudhury:2013dia,Choudhury:2013yg,Kim:1999dq,Lee:2000vf,Kim:2000pz,Kim:2000ym}
 and dilaton coupling \cite{Choudhury:2013eoa,Choudhury:2013dia,Choudhury:2013yg,Choudhury:2013aqa}
 which originates from heterotic string theory.
\item The background warped metric has a RS like structure \cite{Randall:1999ee,Randall:1999vf} on a slice of ${\bf AdS_{5}}$ geometry.
      For example,
     from  10-dimensional string model compactified on ${\bf AdS_5\times S^5}$, one typically obtains
     moduli from ${\bf S^5}$ as scalar degrees of freedom. Such moduli can be stabilized by fluxes.
     In our model, which is similar to a 5-dimensional Randall-Sundrum (RS) model, it is assumed that these 
     degrees of freedom are frozen to their Vacuum expectation value (VEV) and are non-dynamical at the energy scale under consideration \cite{Reece:2010xj}.
     We therefore focus into the slice of ${\bf AdS_5}$ as is done for the 5-dimensional RS model.
\item  The well known ${\bf S^{1}/Z_{2}}$ orbifold compactification is considered.     
\item The dilaton degrees of freedom is assumed to be confined within the bulk.
\item We allow the interaction between dilaton and the 5D bulk cosmological constant via dilaton coupling.
\item The Higgs field is localized at the visible (TeV) brane and the hierarchy problem is resolved via Planck to TeV scale warping.
\item The modulus can be stabilized by introducing
scalar in the ${\bf AdS_{5}}$ bulk without any fine tuning following Goldberger-Wise (GW) mechanism \cite{Goldberger:1999uk,Goldberger:1999un,Goldberger:1999wh,Choudhury:2014hna}.
\item Additionally while determining the values of the model parameters we require
     that the bulk curvature to be less than the five dimensional Planck scale $M_{5}$ so that
     the classical solution of the 5-dimensional gravitational equations can be trusted \cite{Davoudiasl:1999jd,Das:2013lqa}.
\end{itemize}
\subsection{The background setup}
\label{l1a}
Before going to discuss the various features of fermion localization, we start with the 5D action for the two brane
warped geometry model including higher curvature gravity as \cite{Choudhury:2013yg}:
\be\begin{array}{llll}\label{eq1}
 \displaystyle S=\int d^{5}x \left[\sqrt{-g_{(5)}}\left\{\frac{M^{3}_{(5)}}{2}R_{(5)}+\frac{\alpha_{(5)}M_{(5)}}{2}
\left[R^{ABCD(5)}R^{(5)}_{ABCD}-4R^{AB(5)}R^{(5)}_{AB}+R^{2}_{(5)}\right]\right.\right.\\ \left.\left.
\displaystyle ~~~~~~~~~~~~~~~~~~~~~~~~~~~~~~
-M^{3}_{(5)}\frac{g^{AB}}{2}\partial_{A}\chi(y)\partial_{B}\chi(y)-2\Lambda_{5}e^{\chi(y)}\right\}-
\displaystyle \sum^{2}_{i=1}\sqrt{-g^{(i)}_{(5)}}T_{i}e^{\chi(y)}\delta(y-y_{i})\right]
\end{array}\ee
where $A,B,C,D=0,1,2,3,4$. Here $i$ signifies the brane index, $i=1(\text{hidden})$, $2(\text{visible})$
and $T_{i}$ is the brane tension. Additionally $\alpha_{5}$ and $\chi(y)$ represent the 
Gauss-Bonnet coupling and dilaton.
The background metric describing slice of the ${\bf AdS_{5}}$ is given by,
\be\begin{array}{llllll}\label{eq2}
   \displaystyle ds^{2}_{5}=g_{AB}dx^{A}dx^{B}=e^{-2k_{\bf M}(y)r_{c}|y|}\eta_{\alpha\beta}dx^{\alpha}dx^{\beta}+r^{2}_{c}dy^{2}
   \end{array}\ee
where $r_{c}$ represents the compactification radius of extra dimension which has a unit of $M^{-2}_{pl}$.
Here the orbifold points are $y_{i}=[0,\pi]$
and periodic boundary condition is imposed in the closed interval $-\pi\leq y\leq\pi$. After orbifolding, the size of the extra dimensional
interval is $\pi r_{c}$. Here it is important to note that, the bulk extra dimension $y$ is dimensionless and plays a role of an angular coordinate in this conetext. Moreover in the above metric ansatz
$e^{-2A(y)}$ represents the warp factor while  $\eta_{\alpha\beta}=(-1,+1,+1,+1)$ is flat Minkowski metric.
A more general brane metric for a purely Einsteinian bulk has been discussed in \cite{Koley:2010za}.
\subsection{Warp factor from Gauss-Bonnet dilaton gravity}
\label{l1b}
After varying the model action stated in Eq~(\ref{eq1}) we get:
\be\begin{array}{lllll}\label{eq3}
    \delta S=\displaystyle \int d^5x\left[\sqrt{-g_{(5)}}\left\{M^3_{(5)}G^{(5)}_{AB}+\alpha_{(5)}M_{(5)}H^{(5)}_{AB}+T_{AB}\right\}+
\sum^{2}_{i=1}T_{i}\sqrt{-g^{(i)}_{(5)}}g^{(i)}_{\alpha\beta}\delta^{\alpha}_{A}\delta^{\beta}_{B}e^{\chi(y)}\delta(y-y_{i})\right]\delta g^{AB}\\
\displaystyle~~~~~~~~~~~~~~~+ \int d^5x\left[\sqrt{-g_{(5)}}\left\{-M^3_{(5)}\Box_{(5)}\chi-2\Lambda_{(5)}e^{\chi(y)}\right\}+
\sum^{2}_{i=1}T_{i}\sqrt{-g^{(i)}_{(5)}}e^{\chi(y)}\delta(y-y_{i})\right]\delta\chi
   \end{array}\ee
where the five dimensional Einstein's tensor,  the Gauss-Bonnet tensor are given by:
\be\begin{array}{llll}\label{eq4}
    G^{(5)}_{AB}=\left[R^{(5)}_{AB}-\frac{1}{2}g^{(5)}_{AB}R_{(5)}\right],
   \end{array}\ee
and
\be\begin{array}{llll}\label{eq5}
  H^{(5)}_{AB}=2R^{(5)}_{ACDE}R_{B}^{CDE(5)}-4R_{ACBD}^{(5)}R^{CD(5)}
-4R_{AC}^{(5)}R_{B}^{C(5)}+2R^{(5)}R_{AB}^{(5)}\\ ~~~~~~~~~~~~~~~~~~~~~~~~~~~~~~~~~~~~~~~~~~~-\frac{1}{2}g^{(5)}_{AB}
\left(R^{ABCD(5)}R^{(5)}_{ABCD}-4R^{AB(5)}R^{(5)}_{AB}+R^{2}_{(5)}\right).
   \end{array}\ee 
Also the 5D D'Alembertian operator is defined as, 
$\Box_{(5)}\chi(y)=\frac{1}{\sqrt{-g_{(5)}}}\partial_{A}\left(\sqrt{-g_{(5)}}\partial^{A}\chi(y)\right)$.  
In this context the 5D dilaton stress tensor is defined as, 
$T_{AB}=-\frac{2}{\sqrt{-g_{(5)}}}\frac{\delta}{\delta g^{AB}}\left(\sqrt{-g_{(5)}}{\cal L}_{\chi}\right)$,
where ${\cal L}_{\chi}$ is dilaton Lagrangian as given by, 
${\cal L}_{\chi}=\left[-M^3_{(5)}\frac{g^{AB}}{2}\partial_{A}\chi\partial_{B}\chi-V(\chi)\right]$.
Here it is important to note that the 5D bulk %LOCALIZED TERM HAS BEEN OMITTED% 
 dilaton potential is identified to be the following expression:
$V(\chi)=2\Lambda_{5}e^{\chi(y)}$,
where 5D cosmological constant $\Lambda_{(5)}$ fix the scale of the dilaton potential.

By doing explicit computation one can show that the 5D dilaton stress tensor can be expressed as:
\be T_{AB}=\left[\partial_{A}\chi(y)\partial_{B}\chi(y)+g_{AB}\left(\frac{g_{CD}}{2}\partial_{C}\chi(y)\partial_{D}\chi(y)+V(\chi)\right)\right].\ee
After varying the model action stated in equation(\ref{eq1}) with respect to the 5D metric $g_{AB}$ we get, 
$\frac{\delta S}{\delta g^{AB}}=0$ 
using which the 5D bulk equation of motion turns out to be,
\be\begin{array}{lllll}\label{eq31}
    \displaystyle \sqrt{-g_{(5)}}\left[G^{(5)}_{AB}+\frac{\alpha_{(5)}}{M^{2}_{(5)}}H^{(5)}_{AB}\right]
=-\frac{1}{M^{3}_{(5)}}\left[\sqrt{-g_{(5)}}~T_{AB}+
\sum^{2}_{i=1}T_{i}\sqrt{-g^{(i)}_{(5)}}g^{(i)}_{\alpha\beta}\delta^{\alpha}_{A}\delta^{\beta}_{B}e^{\chi(y)}\delta(y-y_{i})\right].
   \end{array}\ee
   Now very far from the orbifold points the 5D field equation in the warped background can be simplified in the following form:
   \be\begin{array}{lllll}\label{eq31c}
       \displaystyle \sqrt{-g_{(5)}}\left[G^{(5)}_{AB}+\frac{\alpha_{(5)}}{M^{2}_{(5)}}H^{(5)}_{AB}\right]
   =-\frac{1}{M^{3}_{(5)}}\sqrt{-g_{(5)}}~T_{AB}.
      \end{array}\ee
 Similarly varying equation(\ref{eq1}) with respect to the dilaton field we get, 
   $\frac{\delta S}{\delta \chi}=0$.
  using which the Klein Gordon (KG) equation for the dilaton in the warped background turns out to be:
\be\begin{array}{llll}\label{eq61}
\displaystyle \frac{1}{M^{3}_{(5)}}\sum^{2}_{i=1}T_{i}\sqrt{-g^{(i)}_{(5)}}e^{\chi(y)}\delta(y-y_{i})
=\sqrt{-g_{(5)}}\left\{
\displaystyle 
2\frac{\Lambda_{(5)}}{M^{3}_{(5)}}e^{\chi(y)}+\Box_{(5)} \chi\right\}.
   \end{array}\ee
  Now very far from the orbifold points the KG equation for the dilaton in the warped background can be simplified in the following form:
  \be\begin{array}{llll}\label{eq611}
  \displaystyle \Box_{(5)} \chi=J(\chi).
     \end{array}\ee
     Here $J(\chi)$ is identified to be the source function for the bulk localized dilaton field, which is given by the following expression:
     \be J(\chi)=\frac{V(\chi)}{M^{3}_{(5)}}= 2\frac{\Lambda_{(5)}}{M^{3}_{(5)}}e^{\chi(y)}.\ee
     Before solving the Eq~(\ref{eq31}) and Eq~(\ref{eq611}), here it is important to note that the following crucial facts which are surely be helpful for us to understand the nature of the field equations:
     \begin{itemize}
     \item First of all the derived two field equations are second order coupled differential equations. Here such complicated structures are appearing due to minimal interaction between gravity and dilaton in the bulk.
     
     \item On the other hand, the KG equation for the dilaton in the warped background is itself complicated as it contains an exponential source term, which is appearing as dilaton effective potential in the bulk.
     
     \item Both of the equations can be simplified as Eq~(\ref{eq31c}) and Eq~(\ref{eq611}) if we go very far drom the orbifold fixed points.
     \end{itemize} 
     In order to solve these coupled equations, we use some approximations as follows :
     %Due to the complicated coupled structures we use the following ansatzs to solve both of the field equations:
     First of all, it is important to note that we may neglect the derivative terms in the dilaton action with respect to it's potential which is of the order of 
     Planck scale. Due to this fact the stress energy tensor of the dilaton approximately, is given by the expression:
     \be T_{AB}\approx g_{AB}V(\chi)=2g_{AB}\Lambda_{(5)}e^{\chi(y)}.\ee
     Consequently, the 5D field equation can be be recast into the following simplified form:
     \be\begin{array}{lllll}\label{eq3c}
     	\displaystyle \sqrt{-g_{(5)}}\left[G^{(5)}_{AB}+\frac{\alpha_{(5)}}{M^{2}_{(5)}}H^{(5)}_{AB}\right]
     	\approx-\frac{e^{\chi(y)}}{M^{3}_{(5)}}\left[\Lambda_{(5)} \sqrt{-g_{(5)}}g^{(5)}_{AB}+
     	\sum^{2}_{i=1}T_{i}\sqrt{-g^{(i)}_{(5)}}g^{(i)}_{\alpha\beta}\delta^{\alpha}_{A}\delta^{\beta}_{B}\delta(y-y_{i})\right],
     \end{array}\ee
     which can be further simplified in the region far away from the orbifold fixed points as:
     \be\begin{array}{lllll}\label{eq31cx}
     	\displaystyle \sqrt{-g_{(5)}}\left[G^{(5)}_{AB}+\frac{\alpha_{(5)}}{M^{2}_{(5)}}H^{(5)}_{AB}\right]
     	\approx-\frac{e^{\chi(y)}}{M^{3}_{(5)}}\sqrt{-g_{(5)}}~\Lambda_{(5)}g^{(5)}_{AB}=-\frac{1}{M^{3}_{(5)}}\sqrt{-g_{(5)}}~V(\chi)g^{(5)}_{AB}.
     \end{array}\ee
     Moreover, considering the equation of motion for the dilaton , we further note that the source term in the right hand side is extremely suppressed due to the warping
     and therefore may be neglected. As a result the space variation of $\chi(y)$ can be derived from the equation: 
     \be\begin{array}{llll}\label{eq611ccxx}
     	\displaystyle \partial^2_{y}\chi(y)\approx 0.
     \end{array}\ee

Now using the ${\bf Z_{2}}$ orbifolding, we obtain 
at the leading order of $\alpha_{(5)}$  \cite{Choudhury:2013yg}:
\be\begin{array}{llll}\label{gradilatonic}
 \displaystyle \boxed{ \chi(y)=\left(c_{1}|y|+c_{2} \right) }
   \end{array}\ee
where $c_{1}$ and $c_{2}$ are arbitrary integration constants in which $c_{1}$ characterizes the strength of the dilaton self interaction within the bulk.
For our computation we fix $c_2 =0$. As the nature of warping influences the localization
profile of the bulk fermion, therefore it is expected that the dilaton charge $c_1$ for a given fermionic field 
will determine the localization property and hence the effective fermion mass term on the brane. 

The corresponding warp factor turns out to be \cite{Choudhury:2013yg}:
\be\begin{array}{llll}\label{warp}
   \displaystyle A(y):= A_{\pm}(y)=k_{\pm}(y)r_{c}|y|\end{array}\ee
where 
\be\begin{array}{llll}\label{wsol}
\displaystyle \boxed{ k_{\pm}(y)=\sqrt{\frac{3M^{2}_{(5)}}{16\alpha_{(5)}}
\left[1\pm\sqrt{\left(1+\frac{4\alpha_{(5)}\Lambda_{5}e^{\chi(y)}}{9M^{5}_{(5)}}\right)}\right]}}~.\end{array}\ee
Also the localized brane tensions are given by:
   \bea \boxed{T_{2}=-T_{1}= 24k_{\pm}(y)M^3_{(5)}~e^{-\chi(y)}\left[1-\frac{\alpha_{(5)}}{3M^2_{(5)}}k^{2}_{\pm}(y)r^2_c\right]}~.\eea
   In the small $\alpha_{(5)}$, $c_{1}$ and $c_{2}$ limit we retrieve the results as in the case of RS model with:
\be \label{rs} 
k_{-}(y)\rightarrow k_{RS}=\sqrt{-\frac{\Lambda_{5}}{24M^{3}_{(5)}}}.\ee 
and the corresponding brane tension is given by:
 \bea T^{RS}_{2}=-T^{RS}_{1}=24k_{RS}M^3_{(5)}.\eea
Here we have discarded 
the +ve branch of solution of $k_{+}$ which diverges in the small $\alpha_{(5)}$ limit, bringing 
in ghost fields \cite{Rizzo:2004rq,Dotti:2007az,Torii:2005xu,Konoplya:2010vz,Kim:2000ym,Nojiri:2010wj}. 
Now expanding Eq~(\ref{wsol}) in the perturbation series order by order around 
$\alpha_{5}\rightarrow 0$, $c_{1}\rightarrow 0$ and  $c_{2}\rightarrow 0$ we can write:
\begin{eqnarray}\label{wsol1}
\displaystyle \boxed{k_{\bf M}(y):=k_{-}(y)=k_{RS}~e^{\frac{\chi(y)}{2}}\left[1+{\bf L}
+{\cal O}({\bf L}^2)+\cdots\right]}~.
\end{eqnarray}
where ${\bf L}$ is defined as:
\be {\bf L}:=\frac{4\alpha_{(5)} k^{2}_{RS}}{M^{2}_{(5)}}.\ee 
However, the results of this paper will be unchnaged if we take the non zero value of the constant $c_2$. In the weak coupling regime of gravity and dilaton one can also consider non zero but small values of $c_2$ which is finally appearing as an overall factor $e^{c_2/2}$ in the expression for the warp factor in Eq~(\ref{wsol1}). More precisely the contribution in the warp factor can be written as:
\bea e^{c_2/2}\approx 1+\underbrace{\left(\frac{c_2}{2}+\cdots\right)}_{\bf <<1~in ~weak~coupling}\approx\left(1+\frac{c_2}{2}\right),\eea
where we have neglected all the higher powers of $c_2$ as in the weak coupling regime of the gravity and dilaton always $c_2<<1$ approximation holds good. On the other hand it is important to note that, in this context strictly one cannot consider very large values of $c_2$ as in that case the perturbative solution in the weak coupling approximation itself is not valid for dilaton and consequently in the solution for the warp factor. Additionally in the weak coupling regime of gravity and dilaton the dilatonic charge $c_1$ is small compared to unity at the orbifold point $y=\pi$ where the visible brane is placed. Here to visualize the effect of dilaton in the phenomena of localization of fermions we further use an approximation that the contribution from the dilatonic charge $c_1$ is larger than the contribution from $c_2$ at the orbifold point $y=\pi$. For the similar reason here also large values of $c_1$ is not strictly allowed.
Using these set of approximations one can finally write down the warp function as:
 \begin{eqnarray}\label{wxsol1}
\displaystyle k_{\bf M}(y)\approx k_{RS}~e^{\frac{c_1y}{2}}\left(1+\frac{c_2}{2}\right)\left[1+{\bf L}
+{\cal O}({\bf L}^2)+\cdots\right].
\end{eqnarray}
 Further Eq~(\ref{wxsol1}) can be used to solve the naturalness or gauge hierarchy problem and for this purpose one need to consider the following modified constraint at the orbifold point $y=\pi$ as given by:
\bea k_{\bf M}(\pi)r_c\approx k_{RS}r_c e^{c_1\pi/2}\left(1+\frac{c_2}{2}\right)\left[1+{\bf L}
+{\cal O}({\bf L}^2)+\cdots\right]\approx 12.\eea
As in the weak coupling regime the constant $c_2$ is just playing the role of a correction term or an overall normalization factor, one can neglect the contribution from the coupling $c_2$ completely without taking care of any small contributions from the small corrections. This is exactly equivalent to the similar effect if we set $c_2=0$ from the strating point of our computation. In this case Eq~(\ref{wxsol1}) can be written as:
\begin{eqnarray}\label{wxdsol1}
\displaystyle \boxed{k_{\bf M}(y)\approx k_{RS}~e^{\frac{c_1y}{2}}\left[1+{\bf L}
+{\cal O}({\bf L}^2)+\cdots\right]}~,
\end{eqnarray}
and consequently the modified constraint condition to solve the naturalness or gauge hierarchy problem can be recast at the orbifold point $y=\pi$ as:
\bea \boxed{k_{\bf M}(\pi)r_c\approx k_{RS}r_c e^{c_1\pi/2}\left[1+{\bf L}
+{\cal O}({\bf L}^2)+\cdots\right]\approx 12}~.\eea
Now if we set the limit $\alpha_{5}\rightarrow 0$ and $c_{1}\rightarrow 0$ then one can get back the RS result, which is $k_{RS}r_c\approx 12$.
Here it is important to note that, both the solution of the warp factor and the dilaton is consistent with Eq~(\ref{eq31cx}) and Eq~(\ref{eq611ccxx}) in the weak coupling regime of gravity (graviton) and dilaton in the bulk.

\subsection{Localization scenario for fermions}
We will now start our discussion regarding the localization scenario of fermionic modes.
The five dimensional action for the massive fermionic field can be written as:
%\begin{widetext}
\be\begin{array}{llll}\label{fer}
 \displaystyle S_{f}=\int d^{5}x\left[Det({\cal V})\right]~\left\{i\bar{\Psi}(x,y)\gamma^{\alpha}
{\cal V}_{\alpha}^{M}\overleftrightarrow{{\large\bf D}_{\mu}}\Psi(x,y)\delta^{\mu}_{M}
%\right.\\ \left.~~~~~~~~~~~~~~~~~~~~\displaystyle 
-sgn(y)m_{f}\bar{\Psi}(x,y)\Psi(x,y)+~h.c.\right\}\\
\displaystyle~~~=\int d^{5}x~e^{- 4k_{\bf M}(y)r_{c}|y|}  \left\{\bar{\Psi}(x,y)
\left[i e^{k_{\bf M}(y)r_{c}|y|} \gamma^{\mu}\partial_{\mu}\right.\right. \\ \left.\left.\displaystyle~~~~~~~~~~~~~~~~~~~~~~~~~~~~~~~~~
 + \gamma^{5} (\partial_{y} -2r_c \partial_{y}\{k_{\bf M}(y)|y|\}) - sgn(y)m_B \right ]{\Psi}(x,y)+h.c.\right\}
   \end{array}\ee
%\end{widetext}
where the differential operator $\overleftrightarrow{{\large\bf D}_{\mu}}$ ise defined as, 
$\overleftrightarrow{{\large\bf D}_{\mu}}:=\left(\overleftrightarrow{\partial_{\mu}}
+\Omega_{\mu}\right)$, which represents the covariant derivative in presence
of fermionic spin connection: 
\be \Omega_{\mu}=\frac{1}{8}\omega_{\mu}^{\hat{A}\hat{B}}\left[\Gamma_{\hat{A}},\Gamma_{\hat{B}}\right].\ee
Here $\omega_{\mu}^{\hat{A}\hat{B}}$ represents the gauge field respecting ${\cal SO}(3,1)$ transformation on the vierbein
coordinate. Here we assume that the bulk fermion mass $m_B$ originates through an underlying spontaneous
symmetry breaking in bulk via 5-dimensional Higgs mechanism \cite{Dey:2009gf,Huber:2000fh,Chang:1999nh,Chang:2000zh,Davoudiasl:2005uu}.
The 5D Gamma matrices: \be \Gamma^{\hat{A}}=\left(\gamma^{\mu},\gamma_{5}:=
\frac{i}{4!}{\bf \epsilon}_{\mu\nu\alpha\beta}\gamma^{\mu}\gamma^{\nu}\gamma^{\alpha}\gamma^{\beta}=i\gamma_{4}\right)\ee satisfy
the Clifford algebra
anti-commutation relation $ \{\Gamma^{\hat{A}},\Gamma^{\hat{B}}\}=2\eta^{\hat{A}\hat{B}}$ with $\eta^{\hat{A}\hat{B}}=$\\$diag\left(-1,+1,+1,+1,+1\right)$.
In this context \be g_{MN}:=\left({\cal V}_{M}^{\hat{A}}\otimes{\cal V}_{N}^{\hat{B}}\right)\eta_{\hat{A}\hat{B}},\ee
where ${\cal V}_{M}^{\hat{A}}$ are characterized by the usual conditions:
\begin{eqnarray}\label{vbcx}
    \displaystyle {\cal V}_{4}^{4}&=&1,\\ 
    \label{vbcx2}{\cal V}_{\mu}^{\hat{A}}&=&e^{k_{\bf M}(y)r_{c}|y|}\delta_{\mu}^{\hat{A}},\\
    \label{vbcx3} Det({\cal V})&=&e^{-4k_{\bf M}(y)r_{c}|y|}.
   \end{eqnarray}
and $\hat{A},\hat{B}$ being tangent space indices. 
For our set up ${\cal SO}(3,1)$ spin connection can be written as: 
\begin{eqnarray}\Omega_{4}&=&0,\\ 
\Omega_{\mu}&=&-\frac{1}{2}e^{-k_{\bf M}(y)r_{c}|y|}k_{\bf M}(y)r_{c}\gamma_{5}\gamma_{\mu}.\end{eqnarray}
Here focusing our attention only to the lowest fermionic mode in the brane. To serve this purpose 
we decompose the five-dimensional spinor as \be \Psi(x,y)=\psi(x) \xi(y).\ee In the 
massless case the definite chiral states $\psi_L(x)$ and $\psi_R(x)$  
correspond to left and right chiral states in four dimension. 
The $\psi_L$ and $\psi_R$ are constructed by, \be \psi_{L,R} = \frac{1}{2} (1 \mp \gamma^5) \psi.\ee 
Here $\xi$ denotes the extra dimensional component of the fermion wave function. 
We then can decompose five-dimensional spinor in the following way:
\be
 \Psi(x,y)=\psi_L (x) \xi_L(y) + \psi_R (x) \xi_R(y)
\ee
Substituting the above decomposition in Eq~(\ref{fer}) we obtain the following equations for the fermions,
%\begin{widetext}
\be\begin{array}{lll}
\label{fermeql}
\displaystyle  e^{-k_{\bf M}(y)r_{c}|y|}\left [\pm (\partial_{y}-2r_c \partial_{y}\{k_{\bf M}(y)|y|\}) %\right. \\ \left.~~~~~~~~~~\displaystyle 
+ sgn(y)m_B \right ] \xi_{R,L}(y)
 = - m ~\xi_{L,R}(y) 
\end{array}\ee
%\end{widetext}
where $m_B$ and $m$ represent the 5D bulk mass and \textcolor{red}{lowest} effective mass of \textcolor{red}{4D} fermion respectively. 
The 4D fermions obey the canonical 
equation of motion, $i \gamma^{\mu} \partial_{\mu} \psi_{L,R} = m  \psi_{L,R}$.
Also it is important note that the left and right handed part of the extra dimensional wave function satisfies the usual ortho-normality condition.
Finally, the solution of Eq~(\ref{fermeql}) turns out to be:
%\begin{widetext}
\be\begin{array}{lll}
\label{sol}
\boxed{\xi_{L,R}(y)={\cal N}\text{exp}\left[2~e^{\frac{\chi(y)}{2}} k_{\bf M}(y)|y|r_{c}
\pm m\int dy~e^{k_{\bf M}(y)r_c |y|}%\right.\\ \left.~~~~~~~~~~~~~~~~~~~~~~~~~~~~~~~~~~~~~~~~~~~~~~~~~~~~ \displaystyle 
\pm sgn(y)m_B |y|\right]}
\end{array}\ee
%\end{widetext}
where ${\cal N}$ represents the normalization constant~\footnote{Applying the normalization of the extra dimensional wave function for left and right chiral fermionic modes the 
normalization constant can be expressed as:
$${\cal N}=\frac{1}{\sqrt{\int^{\pi}_{0}dy~\exp\left[\left(4e^{\frac{\chi(y)}{2}}-3\right)k_{\bf M}(y)r_c |y|\right]}}.$$}.
%%%%%%%%%%%%%%%%%%%%%%%%%%FIGURES%%%%%%%%%%%%%%%%%%%%%%%%%%%%%%%%%

\begin{figure*}
\centering
\subfigure[Left fermionic mode with~$m\sim {\cal O}({\rm TeV})$.]{
    \includegraphics[width=7.5cm,height=5.2cm] {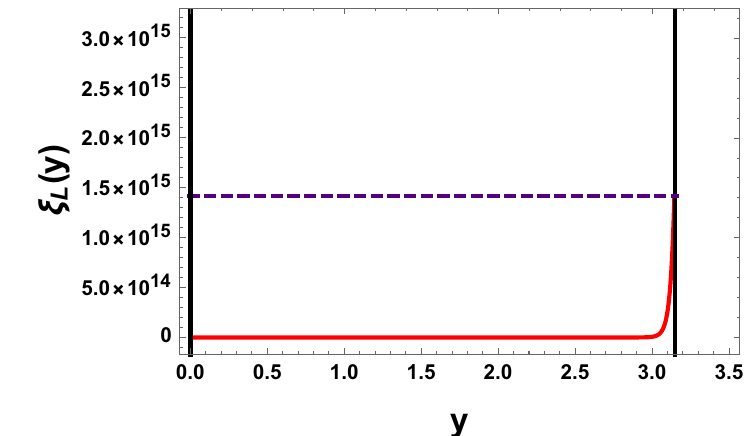}
    \label{x1}
}
\subfigure[Right fermionic mode with~$m\sim -{\cal O}({\rm TeV})$.]{
    \includegraphics[width=7.5cm,height=5.2cm] {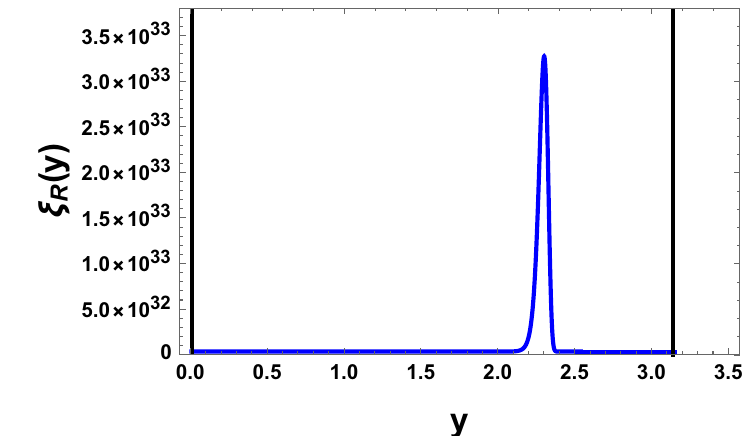}
    \label{x2}
}
\subfigure[Leftt fermionic mode with~$m\sim {\cal O}({\rm GeV})$.]{
    \includegraphics[width=7.5cm,height=5.2cm] {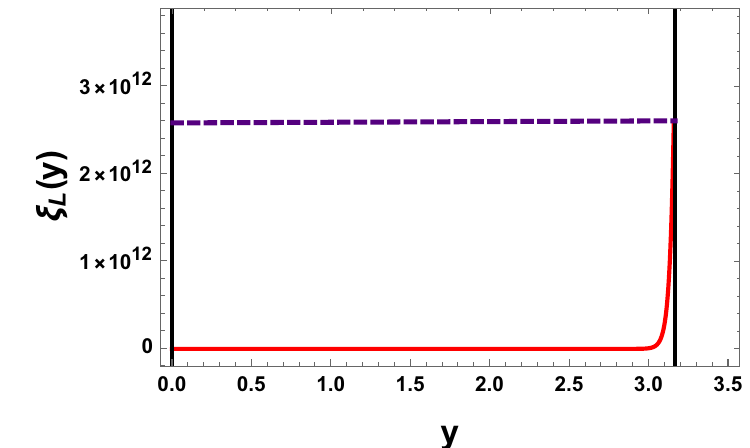}
    \label{x3}
}
\subfigure[Right fermionic mode with~$m\sim -{\cal O}({\rm GeV})$.]{
    \includegraphics[width=7.5cm,height=5.2cm] {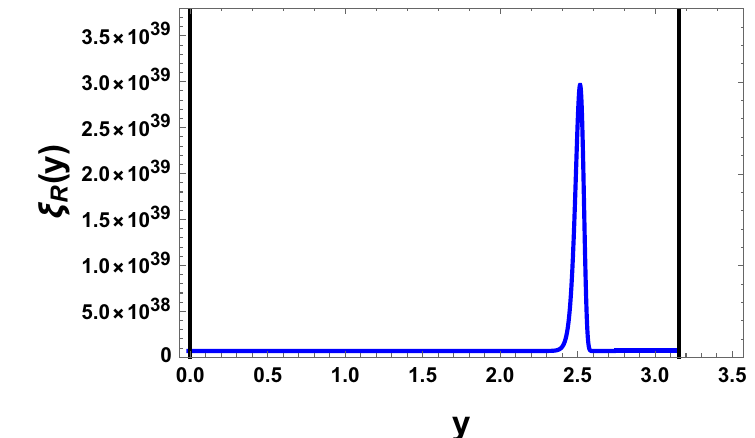}
    \label{x4}
}
\subfigure[Left fermionic mode with~$m\sim {\cal O}({\rm MeV})$.]{
    \includegraphics[width=7.5cm,height=5.2cm] {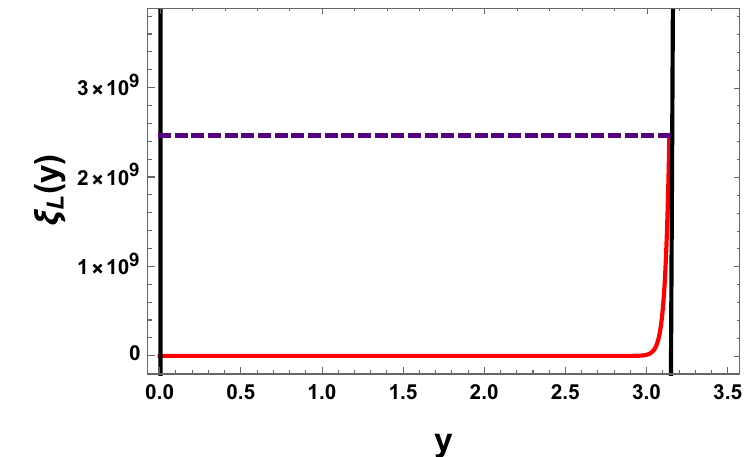}
    \label{x5}
}
\subfigure[Right fermionic mode with~$m\sim -{\cal O}({\rm MeV})$.]{
    \includegraphics[width=7.5cm,height=5.2cm] {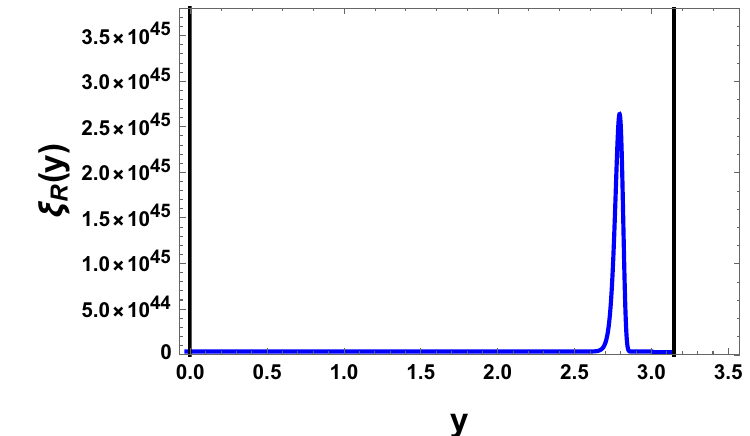}
    \label{x6}
}
%\subfigure[Left fermionic mode with~$m_{n}\sim {\cal O}({\rm keV})$.]{
%    \includegraphics[width=8.5cm,height=2.2cm] {kevl.pdf}
%    \label{x7}
%}
%\subfigure[Right fermionic mode with~$m_{n}\sim -{\cal O}({\rm keV})$.]{
 %   \includegraphics[width=8.5cm,height=2.2cm] {kevr.pdf}
%    \label{x8}
%}
%\subfigure[Left fermionic mode with~$m_{n}\sim {\cal O}({\rm eV})$.]{
%    \includegraphics[width=8.5cm,height=2.2cm] {evl.pdf}
 %   \label{x9}
%}
%\subfigure[Right fermionic mode with~$m_{n}\sim -{\cal O}({\rm eV})$.]{
 %   \includegraphics[width=8.5cm,height=2.2cm] {evr.pdf}
 %   \label{x10}
%}
\caption[Optional caption for list of figures]{ Localization of the left and right handed fermionic profile $\xi_{L,R}(y)$.
For all situations we have taken $\alpha_{(5)}=10^{-3}$,~$c_{1}=0.17$,~$m_{B}=(+1(left),-1(right))$ which fixes, $k_{\bf M}r_{c}=12$, is necessarily required
to solve the hierarchy problem. Here all the masses are given in the Planckian unit. The {\bf black} colored vertical lines represents the hidden and visible branes which are placed at the orbifold points $y=0$ and $y=\pi$ respectively. Also the \textcolor{purple}{\bf purple} colored dashed line represent that left handed mode always touches the visible brane at $y=\pi$ with a finite hight of the wave function.
}
\label{fig1}
\end{figure*}
Fig.~(\ref{x1}-\ref{x6}) describe the localization profiles of the fermion wave function inside the bulk. They clearly depict that
for both massive as well as massless fermions, the left handed mode is localized on the brane while the right handed fermions are localized inside the bulk. 
Additionally, from the prescribed analysis we can observe that the gradual increment in the dilaton coupling $c_{1}$ for a fixed value of Gauss Bonnet parameter ${\bf L}$
within the window~\footnote{Here we choose this specific window for the GB parameter ${\bf L}$ to confront with other bounds on ${\bf L}$ obtained from various astrophysical observations and collider search, which we have mentioned later in Table~(\ref{tab1xz}). Also there is another motivation for choosing the small values of the GB parameter ${\bf L}$ for our present setup is to justify the validity of the perturbation theory in the weak coupling regime of the gravity sector.  }: \be 10^{-3}<{\bf L}<10^{-7},\ee will shift the peak position of the right handed fermionic wave function towards left side of the visible brane towards the bulk. 
In such a situation the hight of the right fermionic mode increases, amount of localization increases and 
the localization position of the 
left fermionic mode slightly shift from the visible brane towards the bulk. Here it is clear from the Fig.~(\ref{x1}-\ref{x6}), the mass of the different generation fermions increases from MeV to TeV, the wave function gets more and more sharply peaked towards the visible brane, implying more localization. We also observe that the 
the effective 4D mass $m$ term decreases as the peak position
of the left handed fermionic mode shifts towards visible brane. Additionally it is important to note that left handed mode always touches the visible brane at $y=\pi$ with a finite hight of the wave function for the different values of the effective 4D mass $m$ lying within {\rm MeV} to {\rm TeV} window. 
%%%%%%%%%%%%%%%%%%%%%%%%%%%%%%%%%%%%%%%%%%%%%%%%%%%%%%%%%%%%%%%%%%%%%%
\begin{table}
\centering
\footnotesize
%\large
\begin{tabular}{|c|c|c|c|}
\hline
\hline
\hline
%Best fit Cosmological parameters
%\hline
  \textcolor{purple}{\bf Serial}& \textcolor{red}{\bf Constraint on}   & \textcolor{blue}{\bf Different sources for}& \textcolor{black}{\bf Final}\\
\textcolor{purple}{\bf number} & \textcolor{red}{\bf GB coupling} &  \textcolor{blue}{\bf the constraint on GB coupling} & \textcolor{black}{\bf remarks}\\
& $({\bf \textcolor{red}{\bf L}})$ &  & \\
 
\hline\hline\hline
\textcolor{purple}{\bf I.} & $0-10^{-7}$ &   Perihelion precession of planetary orbits& Astrophysical \\
& &   and the bending angle of null geodesics.& constraint.\\
\hline 
\textcolor{purple}{\bf II.} &  $10^{-7}-0.2$ &   The disovered Higgs mass & LHC\\
&  & and Higss diphoton and dilepton & constraint. \\
&  &  decays using ATLAS and CMS data.& \\
\hline
\textcolor{purple}{\bf III.} & $(4.8-5.1)\times10^{-7}$ &  Lower bound on the lightest KK graviton mass & Search for \\
&   &   from ATLAS dilepton search. & extra dimensions at LHC.\\
\hline
\textcolor{purple}{\bf IV.} & $<0.25$ &  Positivity of viscosity entropy ratio & AdS/CFT\\
& &  & correspondense.\\
 \hline
\textcolor{purple}{\bf V.} &  $10^{-3}-10^{-7}$ &  Localization of lowest mode of left handed & Constraint from extra \\
& &fermions without using any bulk field. & dimensions and \\
& & & consistent with\\
&  & & \textcolor{purple}{\bf I,II,III,IV}.
\\
\hline
\hline
\hline
\end{tabular}
\vspace{.4cm}
\caption{\label{tab1xz}
 Various constraints on GB coupling from different sources.} 
\end{table}
Additionally it is important to note that, from various experiments and observations following constraints are available for the Gauss Bonnet parameter ${\bf L}\sim \alpha_{(5)}$ (where $k_{RS}\sim M_{(5)}$), which are perfectly consistent with the present window of the the Gauss Bonnet parameter considered in this paper:
\begin{itemize}
\item Astrophysical constraints from the perihelion precession of planetary orbits and the bending angle of null geodesics suggests that the bound on the Gauss Bonnet parameter ${\bf L}\sim \alpha_{(5)}$ lie within the following window \cite{Chakraborty:2012sd}:  
         \be 0< {\bf L}< 10^{-7}.\ee
\item  Collider constraints from the Higgs mass of the resonance
discovered near 125 GeV and the constraints from the $\mu$ parameter for Higss diphoton and dilepton decays using ATLAS \cite{Aad:2012tfa} and CMS \cite{Chatrchyan:2013lba} data within the $5\sigma$ 
 statistical C.L. suggests that the bound on the Gauss Bonnet parameter ${\bf L}\sim \alpha_{(5)}$ lie within the following window \cite{Choudhury:2013eoa}:
 \be 10^{-7}< {\bf L}< 0.2.\ee
 \item  Another phenomenological constraint from the the lower bound on the lightest Kaluza-Klein (KK) graviton mass as obtained from the ATLAS \cite{Aad:2012tfa}
 dilepton search in 7 TeV proton-proton collision suggests that the bound on the Gauss Bonnet parameter ${\bf L}\sim \alpha_{(5)}$ lie within the following window \cite{Choudhury:2013eoa}:
  \be 4.8\times10^{-7}< {\bf L}< 5.1\times 10^{-7}.\ee
\item In the context of AdS/CFT the viscosity entropy ratio can
be computed in presence of Gauss Bonnet parameter ${\bf L}\sim \alpha_{(5)}$ using the well
known Kubo formula as \cite{Choudhury:2013dia,Brigante:2007nu}:
\be \frac{\eta}{S}=\frac{1}{4\pi}(1-4{\bf L})+{\cal O}({\bf L}^2).\ee
To satisfy the constraint on the positivity of the viscosity entropy ratio the upper bound on the Gauss Bonnet parameter ${\bf L}\sim \alpha_{(5)}$ is given by \cite{Choudhury:2013eoa,Brigante:2007nu}:
\be {\bf L}< 0.25.\ee
\end{itemize}
In table~(\ref{tab1xz}) for the comparison of different sources of contraints on GB coupling we have mentioned all of these results, which shows that the result obtained in this paper is perfectly consistent with the astrophysical and particle collider data.

The overlap wave function of the left and right handed mode on the visible brane 
determines the effective mass of the fermion on the 3 brane. 
The effective 4D mass can be computed from the overlap integral as:
 \begin{eqnarray}\label{operlap}
\displaystyle {\bf I}_{overlap}&=&m_B\int d^5x \left[Det({\cal V})\right]~sgn(y)\bar{\Psi}(x,y){\Psi}(x,y)\nonumber\\
\displaystyle &=&\int d^4x ~m_{L,R}\left[\bar{\Psi}_{L}(x){\Psi}_{R}(x)
+\bar{\Psi}_{R}(x){\Psi}_{L}(x)\right],
  \end{eqnarray}
where the 4D effective mass $m_{L,R}$ is given by:
%\begin{widetext}
\be\begin{array}{lllll}
 \displaystyle m_{L,R}= m_{B}\int^{\pi}_{0}dy~e^{-4k_{\bf M}(y)r_c |y|}sgn(y)\xi^{\dagger}_{L}(y)\xi_{R}(y)
\label{d2}.
\end{array}\ee
%\end{widetext}
Above equation clearly indicates that if the bulk mass $m_B=0$ , then the mass of the lowest mode of fermions
will also be zero. Further substituting Eq~(\ref{sol}) in Eq~(\ref{d2}) the effective mass $m=m_{L,R}$ can be recast as:

 \be\begin{array}{llll}\label{operlap1}
\displaystyle \boxed{m=2m_B\frac{\sqrt{5\left[1+{\bf L}
+{\cal O}({\bf L}^2)\right]}\exp\left[\left\{48\pi~e^{\frac{c_{1}\pi}{2}}\left(e^{\frac{c_{1}\pi}{2}}-1\right)+\frac{6}{5c_1}\right\}\left[1+{\bf L}
+{\cal O}({\bf L}^2)\right]\right]}{\sqrt{\frac{\pi}{6c_1}}\left(\text{Erfi}\left[\frac{\sqrt{6}\sqrt{\left[1+{\bf L}
+{\cal O}({\bf L}^2)\right]}}{\sqrt{5c_{1}}}\right]-\text{Erfi}\left[\frac{\sqrt{6}\sqrt{\left[1+{\bf L}
+{\cal O}({\bf L}^2)\right]} \left(1+5 c_1 \pi\right)}{\sqrt{5c_{1}}}\right]\right)}}
  \end{array}\ee
where we fix $k_{RS}r_{c}=12$, which is necessary condition to resolve the gauge hierarchy or naturalness problem. In Eq.~(\ref{operlap1}), we have introduced {\it imaginary error function}, which is defined as:
\bea \text{Erfi}(x)&=&-i~\text{Erfi}(ix)=\frac{2}{\sqrt{\pi}}e^{x^2}~{\cal D}(x)\eea
where ${\cal D}(x)$ is the {\it Dwason function}, is given by:
\bea {\cal D}(x)&=& e^{-x^2}\int^{x}_{t=0}e^{t^2}~dt.\eea
%%%%%%%%%%%%%%%%%%%%%%%%%%%%%%%%%%%%%%%%%%%%%%%%%%%%%%%%%%%%%%%%%%%%%%
\begin{table}
\centering
%\large
\begin{tabular}{|c|c|c|c|}
\hline
\hline
\hline
%Best fit Cosmological parameters
%\hline
 \textcolor{red}{\bf 4D mass}& \textcolor{blue}{\bf 5D bulk}& \textcolor{purple}{\bf GB}  & {\bf Dilaton}  \\
&  \textcolor{blue}{\bf mass}&  \textcolor{purple}{coupling} &  {\bf coupling}  \\
 $\textcolor{red}{\bf m}$ & $\textcolor{blue}{m_{\bf B}}$& $\textcolor{purple}{\bf L}$ &  ${\bf c_1}$\\
(\textcolor{red}{\bf in~ ${\rm GeV}$})& (\textcolor{blue}{\bf in~ $M_{Pl}$})&   & \\
\hline\hline\hline
$10^{3}$ & 1 & $10^{-3}-10^{-7}$ &   0.007\\
\hline
$1$ & 1 & $10^{-3}-10^{-7}$ &  0.033\\
\hline 
$10^{-3}$ &1 & $10^{-3}-10^{-7}$ &  0.057\\
\hline
 $10^{-6}$ & 1 & $10^{-3}-10^{-7}$ &  0.078\\
\hline
$10^{-9}$ & 1 & $10^{-3}-10^{-7}$ &  0.098
\\
\hline
\hline
\hline
\end{tabular}
\vspace{.4cm}
\caption{\label{tab1}
 Parameter space required to generate overlap of left and right handed fermion wave functions at the visible brane via 4D effective mass.} 
\end{table}
In table~(\ref{tab1}) we have shown the total parameter space for the GB coupling ${\bf L}$, dilaton coupling $c_{1}$ and the 5D bulk mass $m_{\bf B}$ required to generate the overlap of left and right handed fermion wave functions at the visible brane via arying the 4D effective mass within the window $10^{-9}~{\rm GeV}<m<10^3~{\rm GeV}$.

\section{Conclusion}
\label{aa6}
 The localization of left handed standard model fermions requires an external 5D bulk 
field. In this work, we have shown that it is possible to localize the SM fermions in the bulk using the higher
curvature dilaton coupled gravity set-up without invoking any external scalar field in the bulk. This, then naturally
explains the origin of localization of left handed fermions in the visible brane whereas the right handed
fermionic modes get delocalized and obtain their peak inside the bulk. We have also obtained the effective
4D mass term in the brane which depends on the GB coupling parameters and dilaton coupling. Thus a string inspired background with higher curvature Gauss-Bonnet term and dilaton field in
the bulk offers a natural explanation for the fermion localization on our brane.

%%%%%%%%%%%%%%%%%%%%%%%%%%%%%%%%%%%%%%%%%%%%%%%%%%%%%%%%%%%%%%%%%%%%%%%%%%%%%%%%%%%%%%%%%%%%%%%%%%%%%%%%%%%%%%%%%%%%%%%%%%%%%%%%%%%%%%%%%%%%%%%%%%%%%%%%%%%%%%%%%%%%%%%%%%%%%%%%%%%%%%%%%%%%%%%%%%%%%%%%%%%%%%%%%%%%%%%%%%%%%
%%%%%%%%%%%%%%%%%%%%%%%%%%%%%%%%%%%%%%%%%%%%%%%%%%%%%%%%%%%%%%%%%%%%%%%%%%%%%%%%%%%%%%%%%%%%%%%%%%%%%%%%%%%%%%%%%%%%%%%%%%%%%%%%%%%%%%%%%%%%%%%%%%%%%%%%%%%%%%%%%%%%%%%%%%%%%%%%%%%%%%%%%%%%%%%%%%%%%%%%
%\newpage
\section*{Acknowledgments}
SC would like to thank Department of Theoretical Physics, Tata Institute of Fundamental Research, Mumbai and specially the Quantum Structure of the Spacetime Group
for providing me Visiting (Post-Doctoral) Research Fellowship. The work of SC was supported
in part by Infosys Endowment for the study of the Quantum Structure of Space Time.  SC and JM thanks Indian Association for the Cultivation of Science (IACS), Kolkata
for various support during the work.  
Last but not the least, we would all like to acknowledge our debt to the people of
India for their generous and steady support for research in natural sciences, especially for theoretical high energy physics, string theory and cosmology.

\end{document}